\documentclass[aps,prb,amsmath,amssymb]{revtex4}
\usepackage{bm}
\usepackage{xfrac}
\usepackage{amsmath}
\usepackage{amssymb}
\usepackage{amsfonts}
\usepackage{graphics,graphicx}
\usepackage{epsf}
\usepackage[%
  colorlinks=true,
  urlcolor=blue,
  linkcolor=blue,
  citecolor=blue
]{hyperref}
\usepackage[all]{hypcap}
\usepackage{color}
\usepackage[utf8]{inputenc}

\newcommand{\be}{\begin{equation}}
\newcommand{\ee}{\end{equation}}
\newcommand{\bea}{\begin{eqnarray}}
\newcommand{\eea}{\end{eqnarray}}
\newcommand{\beb}{\begin{eqnarray*}}
\newcommand{\eeb}{\end{eqnarray*}}

\newcommand{\phrb}[3]{Phys. Rev. B{\bf #1}, #2 (#3).}
\newcommand{\phrl}[3]{Phys. Rev. Lett. {\bf #1}, #2 (#3).}
\begin{document}
\title{Bubble phase at $\nu=1/3$ for a spinless 
hollow-core interaction}

\author{Gr\'egoire Misguich$^{1,2}$, Thierry Jolicoeur$^1$ and Takahiro Mizusaki$^3$}


\affiliation{$^1$~Université Paris-Saclay, CNRS, CEA, Institut de Physique Théorique, 91191 Gif-sur-Yvette, France}
\affiliation{$^2$~Laboratoire de Physique Théorique et Modélisation, CNRS UMR 8089, Université de Cergy-Pontoise, 95302 Cergy-Pontoise, France}
\affiliation{$^3$~Institute of Natural Sciences, Senshu University,  Tokyo 101-8425, Japan}

\date{November, 2020}
\begin{abstract}
We investigate fractional quantum Hall states for model interactions restricted 
to  a repulsive hard-core. When the hard-core excludes relative angular 
momentum 
$m=1$ between spinless electrons the ground state at Landau level filling factor 
$\nu=1/3$ is known to be exactly given by the Laughlin wavefunction. When we exclude 
relative angular momentum three only, W\`ojs, Quinn and Yi have suggested the 
appearance of a liquid state with non-Laughlin correlations. We study this 
special hard-core interaction at filling factor 1/3 on the sphere, 
 torus and cylinder geometry. An analysis of the charged and neutral gaps
on the sphere geometry points to a gapless state. On the torus geometry the 
projected static structure factor has a two-peak feature pointing to 
one-dimensional density ordering. To clarify the nature of the ground state we
perform extended DMRG studies on the cylinder geometry for up to 30 particles. The pair correlation function allows us to conclude that the ground state is a two-particle bubble phase.
\end{abstract}
\pacs{73.43.-f, 73.22.Pr, 73.20.-r}
\maketitle
\section{introduction}
The quantum Hall effect is a striking phenomenon in condensed matter physics. 
It appears as a low-temperature anomaly in the transport properties
of some two-dimensional electronic systems. For special values of an applied 
perpendicular magnetic field the longitudinal resistance goes to zero with an 
activated law
as a function of the temperature and at the same time there is a plateau in the 
Hall resistance.
The one-electron spectrum in these special circumstances consists of  Landau 
levels with macroscopic degeneracy separated by the cyclotron energy. Coulomb
interactions between electrons inside lowest-lying Landau levels give rise to
a family of incompressible liquid states that are responsible for the fractional 
quantum Hall effect (FQHE). Theoretical understanding of the most prominent
state at filling factor $\nu=1/3$ of the lowest Landau level (LLL) is based on an 
explicit many-particle wavefunction due to Laughlin~\cite{Laughlin}. 
The composite fermion (CF) theory of Jain~\cite{JainBook} is also based on 
explicit 
wavefunctions
and capture successfully many physical properties of other FQHE states.
These wavefunctions are however not exact eigenstates of the Coulomb interaction 
Hamiltonian projected onto the LLL. In the case of the Laughlin state it is known that it is the exact 
ground state of a hard-core interaction that gives a nonzero energy only
to states with relative angular momentum $m=1$ between spinless 
electrons~\cite{HaldaneBook}.
The physical relevance of the Laughlin state stems from the fact that one can adiabatically 
follow a path in Hamiltonian space between this special hard-core model and the complete
Coulomb interaction, without closing the gap. No such model is known for the Jain CF wavefunctions.
The CF theory explains the appearance of the experimentally prominent
series of FQHE states observed for filling factors $\nu=p/(2p\pm 1)$ with $p$ a 
positive integer. However this does not exhaust the observed incompressible 
states. For example in the LLL for filling factors between $\nu=1/3$ and 
$\nu=2/5$ two-dimensional electron gases with high mobility also exhibit additional 
fractions at $\nu=5/13,4/11$. There is also a fraction with $even$ denominator
in the second Landau level at $\nu=5/2$ which may very well be the so-called 
Pfaffian state~\cite{Moore91,Pakrousky}. The Laughlin/CF states are built with 
some 
Jastrow-like 
correlation factors giving them the correct low-energy properties. The range 
of validity of these correlations is not yet known. W\`ojs and 
Quinn~\cite{Wojs2000,Wojs2002} have 
argued that the repulsive potential between electrons should have a special
``super-harmonic'' dependence on the relative angular momentum.

If we consider that FQHE states between $\nu=1/3$ and $\nu=2/5$
are due to condensation of quasiparticles or quasiholes emanating from the 
parent state at $\nu=1/3$
then it is not clear what is the effective interaction between the 
quasiparticles/quasiholes. Notably it may be that they are not of the 
Laughlin/CF type. W\`ojs, Yi and Quinn (WYQ) in a series of 
work~\cite{Wojs2004,Wojs2005,Quinn2009,Simion2008} have suggested 
that there is an incompressible state at filling factor $\nu=1/3$
with non-Laughlin correlations. They consider a special hard-core Hamiltonian
which gives nonzero energy only for two-body states with relative angular 
momentum  (RAM) $m=3$.
This may be called a ``hollow-core'' model
since the most repulsive part of the interaction induced by the RAM $m=1$ 
interaction is artificially set to zero.
By use of extensive exact diagonalizations on 
the sphere 
geometry they have given evidence for a series of states that are fluid-like,
e.g. with a ground state with zero total angular momentum, and that have several
hallmarks of the previously known FQHE states. This series appears for a specific relationship 
between the number of electrons $N_e$ and the number of flux quanta through the 
sphere $N_\phi$~: $N_\phi = 3N_e -7$. This is to be contrasted with the series corresponding to the
Laughlin state which happens for $N_\phi = 3N_e -3$. The offset $7$ vs $3$
in the flux-number of particles is called the shift quantum number and is 
related to the topological properties of the state. This state, which does not 
belong to the CF family, has been 
proposed~\cite{Mukherjee2012,Mukherjee2014,Mukherjee2014b,Mukherjee2015,MukherjeeL2015,
Balram2013,Balram2015,Balram2015b,Hutasoit2016} as a candidate to describe some
of the weaker FQHE states at $\nu=4/11,5/13$. No explicit candidate wavefunction is known
up to now for the WYQ state. Detailed studies are required to understand whether it is really
a new type of FQHE or whether it is a state breaking translation symmetry like 
a stripe or a bubble phase~\cite{Koulakov96,Fogler96,Fogler97,Moessner96} which 
are not easily discovered on the sphere geometry.
It may also be related to quantum Hall nematic 
states~\cite{Kivelson98,Fradkin99} as proposed in a recent 
study~\cite{regnault2017}.

In this paper we study the WYQ hollow-core model by exact diagonalizations using 
sphere~\cite{HaldaneBook,Fano86} and torus geometry~\cite{Haldane85}. 
Since the number of particles we can treat is limited to small values
we also use the DMRG algorithm on the cylinder geometry to clarify the nature of the ground state.  For filling factor $\nu=1/3$ we show that the values of gaps extracted 
from charged and neutral excitations extrapolate smoothly to zero in the thermodynamic 
limit from sphere calculations if we stick to the special WYQ shift $N_\phi = 
3N_e -7$. This indication of compressibility alone does not however allow us to understand the nature
of the ground state. We thus use next
 the torus geometry where there is no shift and so there is direct competition
with the Laughlin-like physics as well as competition with states
that break translation invariance like stripe or bubble phases. We find that
the LLL-projected static structure factor of the WYQ state has several peaks
indicating the tendency to spontaneous breakdown of translation symmetry
as is observed in the bubble phase in higher Landau levels.
If one uses a rectangular unit cell one observes the appearance of a special wavevector
where fluctuations are enhanced, indicative of a tendency to spatial ordering.
To clarify the nature of this ordering we
next ran DMRG studies in the cylinder geometry
on large systems of up to 30 electrons to be reasonably sure to avoid finite size effects. 
We measure the pair correlation function
and conclude that the ground state is a two-electron bubble phase as expected
from Hartree-Fock theory for Coulomb interactions in higher Landau levels.

It has been proposed that interactions between CF in the effective second landau level
of the CF theory are modelled after the hollow-core pure $V_3$ model. 
Following this idea it means that an incompressible ground state of the hollow-core model
would explain incompressibility at electronic filling factor $\nu=4/11$ after the standard
Jastrow factor attachment in composite fermion theory since this filling corresponds
to filling $\nu^*=1+1/3$ of the CFs.
However
it is known that the hollow-core model is not always a good model of composite fermions
interactions in the second ``Lambda''-Landau-level as stated in Ref.(\onlinecite{Mukherjee2014}).
Nevertheless CF diagonalization studies~\cite{Mukherjee2014} with the pure Coulomb interactions 
have given evidence for an incompressible state 
 of electrons at filling factor $\nu=4/11$ that happens with an
 unconventional shift (the same as proposed by WYQ) for the CFs. Our results mean that to generate
 a valid incompressible state in this universality class, if it exists, one has to go beyond the simple 
 pure hollow-core model. It remains to be seen whether it is possible to deform the hollow-core model and 
 reach the WYQ universality class.

For filling factor $\nu=1/5$ there is evidence~\cite{Wojs2009} for a series of states with 
$N_\phi = 5N_e - 9$ starting at $N_e=5$ up to $N_e=12$ which are isotropic and have zero angular momentum. The shift is again 
different from that of the Laughlin state at $\nu=1/5$. This special state when expressed in the 
standard Fock space basis has only integer coefficients for all the accessible 
sizes we could reach. This does not happen for the WYQ state which has always 
nonzero energy. The property of integer coefficients is reminiscent of 
the Jack polynomials~\cite{BH1,BH2,BH3,BH4} that describe several special 
states many of them being
critical with zero gap in the thermodynamic limit. In the torus geometry the 
hollow-core Hamiltonian has a set of zero-energy ground states that grows with 
the system size. 

In section II we give the basic formalism for hard-core models of the FQHE. 
Section III is devoted to the study of the thermodynamic limit of the fraction $\nu=1/3$ for 
the hollow-core model on the sphere and on the torus geometry. Section IV
gives our findings from DMRG studies in the cylinder geometry and we present results 
giving evidence for a 2-electron bubble phase. Finally section V presents our conclusions.
The appendices contain some findings related to the special state at $\nu=1/5$.

\section{Hard core models for  fractional quantum Hall states}
In this work
we consider only spin-polarized electronic systems. In the symmetric gauge
defined by the vector potential $\mathbf{A}=(\mathbf{B}\times\mathbf{r})/2$
the LLL basis states can be written as~:
\be
\phi_k(z)=\frac{1}{\sqrt{2^{k+1}\pi}} \, z^k \,\,{\mathrm{e}}^{-|z|^2/4\ell^2},
\label{disk}
\ee
where $k$ is a positive integer which is the disk angular momentum of the state,
$\ell=\sqrt{\hbar/eB}$ is the magnetic length and
$z=x+iy$ is the complex coordinate of the particle.
A generic many-body state for $N_e$ electrons is thus of the form~:
\be
\Psi(z_1,\dots,z_{N_e})= P(z_1,\dots,z_{N_e}) \,\,{\mathrm{e}}^{-\sum_i|z_i|^2/4\ell^2},
\label{manybody}
\ee
where $P$ is an antisymmetric polynomial. Since the exponential factor is 
universal i.e. does not depend on the precise state we will omit it in what 
follows. 
The Laughlin wavefunction is defined as a power of the Vandermonde determinant~:
\be
\Psi^{(p)}_{L}=\Psi_V^m =\prod_{i<j}(z_i-z_j)^p .
\label{LJ}
\ee
It describes successfully the FQHE state at $\nu=1/3$ (resp. $\nu=1/5$) for 
$p=3$ (resp. $p=5$).

A generic two-body interaction Hamiltonian projected onto the LLL can be 
written as a sum of projectors onto states of 
definite relative angular momentum $m$~:
\be
\mathcal{H}=\sum_{i<j}\sum_m V_m \mathcal{{\hat P}}^{(m)}_{ij},
\ee
where $m$ is a non-negative integer  and the coefficients $V_m$ are the 
so-called Haldane pseudopotentials~\cite{HaldaneBook}. The antisymmetry of their wavefunction makes spinless fermions 
sensitive only to odd values of $m$. The set of pseudopotentials 
$\{V_m\}$ thus 
completely characterizes the projected two-body interactions. For the physically 
relevant case of 
the Coulomb interaction the $V_m$ are monotonic and decreasing  with large $m$ 
as 
$\sim m^{-1/2}$.  If we consider the hard-core 
Hamiltonian $\mathcal{H}_1$
with $V_1=1$ and all other pseudopotentials set to zero $V_m=0,m>1$ then
$\mathcal{H}_1$ has many zero-energy eigenstates but the densest such state
corresponding to a polynomial of smallest total degree is unique and is given 
precisely by the Laughlin wavefunction for $p=3$~: $\Psi^{(3)}_{L}$. Similarly 
we can construct
an Hamiltonian with $\Psi^{(5)}_{L}$ as its exact densest zero-energy state
by taking $\mathcal{H}_1+\mathcal{H}_3$ where $\mathcal{H}_3$ has only
$V_3$ nonzero positive pseudopotential. In fact any linear combination with positive 
coefficients of $\mathcal{H}_1$ and $\mathcal{H}_3$ has this property.
The pseudopotentials offer a convenient way to parametrize deformations of two-body Hamiltonians.
For example the belief that the Laughlin state captures correctly the physics of the Coulomb
interacting electrons at filling factor $\nu=1/3$ is based on numerical studies that tune
pseudopotentials from the pure hard-core $V_1$ model to the Coulomb values.
The model we focus on in the paper has only $V_3$ nonzero and will be called the hollow-core model in what follows.

\section{The Wojs-Yi-Quinn series $N_\phi=3N_e-7$}

\subsection{Sphere study}
\begin{figure}[t]
\centering
 \includegraphics[width=0.6\columnwidth]{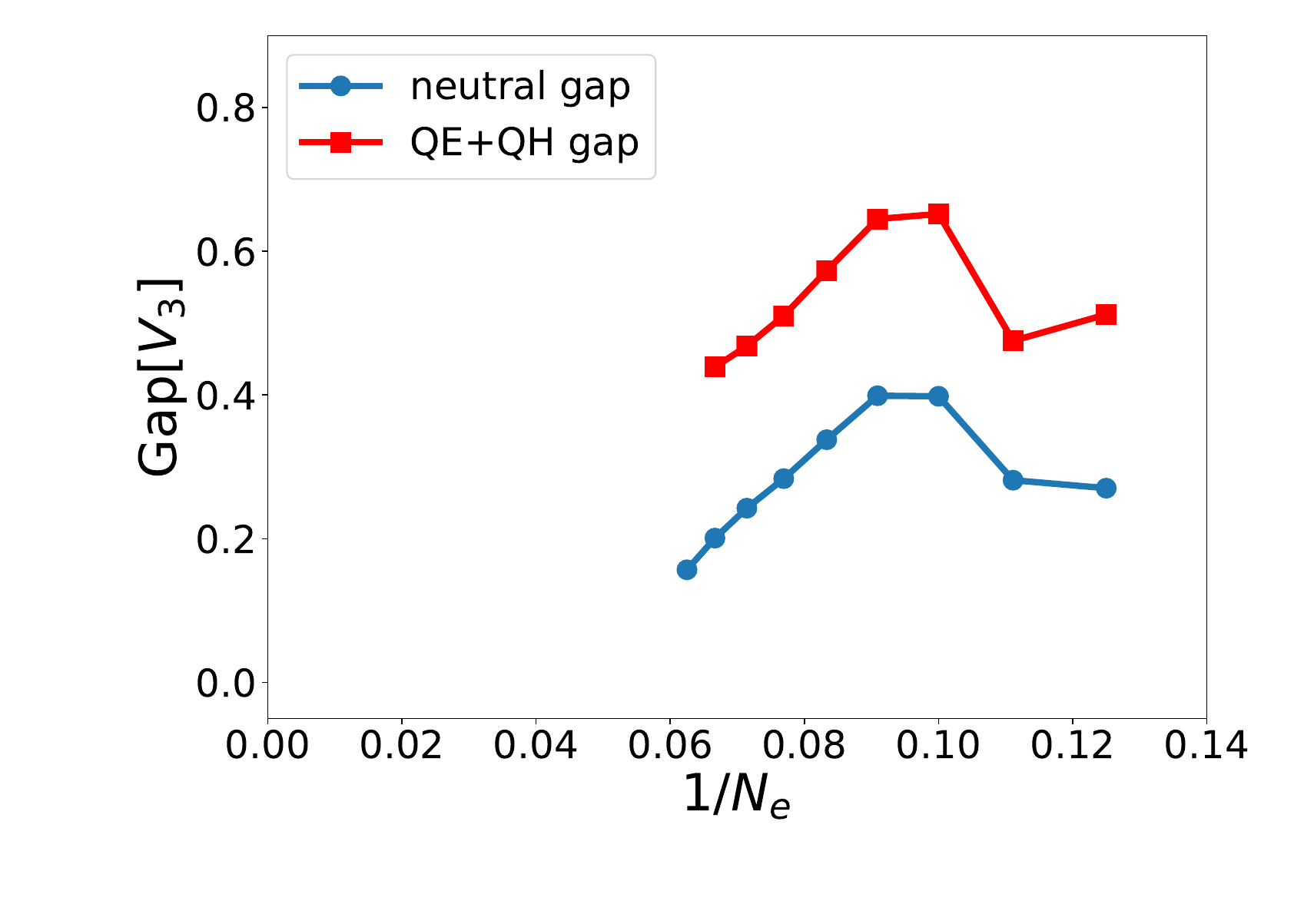}
 \caption{The gaps for the WYQ sequence of states with $N_\phi=3N_e-7$
vs. inverse number of particles.
Lower graph gives the neutral excitation gaps defined without change of the 
flux and the excited state may have any orbital angular momentum. Upper graph 
gives the quasiparticle-quasihole gap defined through addition/removal of one 
flux quantum. Sizes range from $N_e=8$ to $16$ in the the neutral case and up 
to $N_e=15$ in the charged case.}
 \label{GSP}
\end{figure}

In this geometry,
electrons are constrained to 
move 
at the surface of a sphere of radius $R=\ell\sqrt{S}$ with $S=N_\phi/2$ and the LLL 
basis states are given by~:
\be
\Phi_M^{(S)}=
\sqrt{\frac{2S+1}{4\pi}\binom{2S}{S+M}}
u^{S+M}\,v^{S-M},
\quad M=-S,\dots,+S.
\label{basis}
\ee
where  $M$ is a half-integer and we have introduced the elementary spinors~:
\be
u=\cos(\theta/2)\, {\mathrm{e}}^{i\gamma/2},\quad 
v=\sin(\theta/2)\, {\mathrm{e}}^{-i\gamma/2},
\ee
where $\theta,\gamma$ are spherical coordinates.
The basis states form a multiplet of angular momentum $L_{orb}=S$. In this geometry
the Laughlin wavefunction can be written as~:
\be
\Psi^{(p)}_{L} =\prod_{i<j}(u_i v_i - u_j v_i)^p .
\label{LJsph}
\ee
The wavefunction Eq.~(\ref{LJsph}) is a singlet of zero total 
orbital angular momentum since it involves only combinations of factors
$u_iv_j-u_jv_i$ which are themselves singlets. We use the spherical geometry
in some of our exact diagonalization studies. Hence the eigenstates can be 
classified by their total angular momentum.
On the sphere geometry incompressible FQHE states have distinct characteristic 
features. Notably the ground state is a singlet of total orbital angular 
momentum and there is a gap above this ground state which is large
with respect to the finite-size spacing typical of higher-lying levels. This is 
at least the case for the standard Laughlin state at $\nu=1/3$ in the LLL and 
many other FQHE states. In the case of the $\nu=1/3$ state if we add one flux 
quantum then the ground state becomes an isolated multiplet with $L_{orb}=N_e/2$
which is the quasihole. Similarly the quasielectron state involves removal of 
one flux quantum with respect to the fiducial state following $N_\phi=3(N_e-1)$.
Wojs, Yi and Quinn~\cite{Wojs2004} have shown by exact diagonalizations up to 
12 electrons for the $\mathcal{H}_3$ model i.e. with only nonzero 
pseudopotential $V_3$ that there is also a series of states with essentially 
the same 
spectral signatures as the $\nu=1/3$ FQHE state but with a distinct relation 
between flux and number of electrons given by $N_\phi=3N_e-7$. Even if there is 
no clear collective mode resembling the magnetoroton, the ground state is well 
separated from higher-lying continuum for all accessible sizes.
Taken at face value these results imply the existence of a FQHE state at 
$\nu=1/3$ which is topologically distinct from the Laughlin fluid. However one 
has to check the convergence to the thermodynamic limit. Here we have studied 
the gap of this system as a function of the number of electrons. The first gap 
one can define is the lowest excitation energy at $N_\phi=3N_e-7$ irrespective
of its quantum number. In the standard Laughlin case it is the gap to the 
minimum of the magnetoroton branch. This neutral gap is displayed in the lower 
part Fig.~(\ref{GSP}). One can also define a gap  by~:
\be
\Delta_{N}=E_0(N_\phi+1)+E_0(N_\phi-1)-2E_0(N_\phi),
\label{qhgap}
\ee
where $E_0(N_\phi)$ is the ground state of the system with $N_e$ electrons at 
flux $N_\phi$. This gap, when nonzero in the thermodynamic limit, signals a 
cusp in the energy as a function of density. It is given by the upper curve in 
Fig.~(\ref{GSP}). Assuming creation of quasihole/quasielectron by addition/removal
of one flux quantum, this quantity is the gap for creating one quasielectron-quasihole pair. 

Concerning neutral excitations no amount of fitting can possibly lead to a nonzero
value of the gap in the thermodynamic limit since all data points beyond $N_e=11$
display a downward curvature. Concerning the charged gap Eq.~(\ref{qhgap})
the best linear fit to data beyond $N_e=11$ leads also to a large negative value of 
the gap $\approx -0.144V_3$. The best one can do to obtain a positive value is to exclude
all data points except the two largest systems and make a linear fit leading
to an estimate of $\approx +0.032V_3$ an effect due to the small upward curvature that 
appears for the largest systems. Since this gap is much smaller than the overall best fit, the best
guess is that the charged gap is zero and the state is compressible. Since gap scaling
alone may not be a clear signature of the incompressible behavior we turn to the use
of DMRG in section IV to access much larger systems. Note that we have not used any rescaling of the
magnetic length because this is valid only in the case of Coulomb potential. Indeed when the 
gap is known to scale as the inverse of the magnetic length it is sensible to correct for
the fact that the non-zero shift in the flux-number of particles relation changes the density
with respect to thermodynamic limit. However in the case of hard-core models the gap is not
proportional to a simple power of $\ell$ because there is no power-law expression of the
potential in real-space.

We note that a scenario in which the charge gap is nonzero but the neutral gap is vanishing
corresponds to a quantum Hall nematic phase (see e.g. ref.(\onlinecite{BoYang2020}) and references 
therein). This special state of matter breaks rotation symmetry but respects translation 
invariance. General microscopic conditions~\cite{BoYang2020} suggest that one may have to add some extra
hard-core components to the Hamiltonian like a $V_5$ pseudopotential to stabilize
it. We will use the pair correlation function in section (\ref{dmrgres}) to constrain such
a possibility.

\begin{figure}[t]
\centering
 \includegraphics[width=0.5\columnwidth]{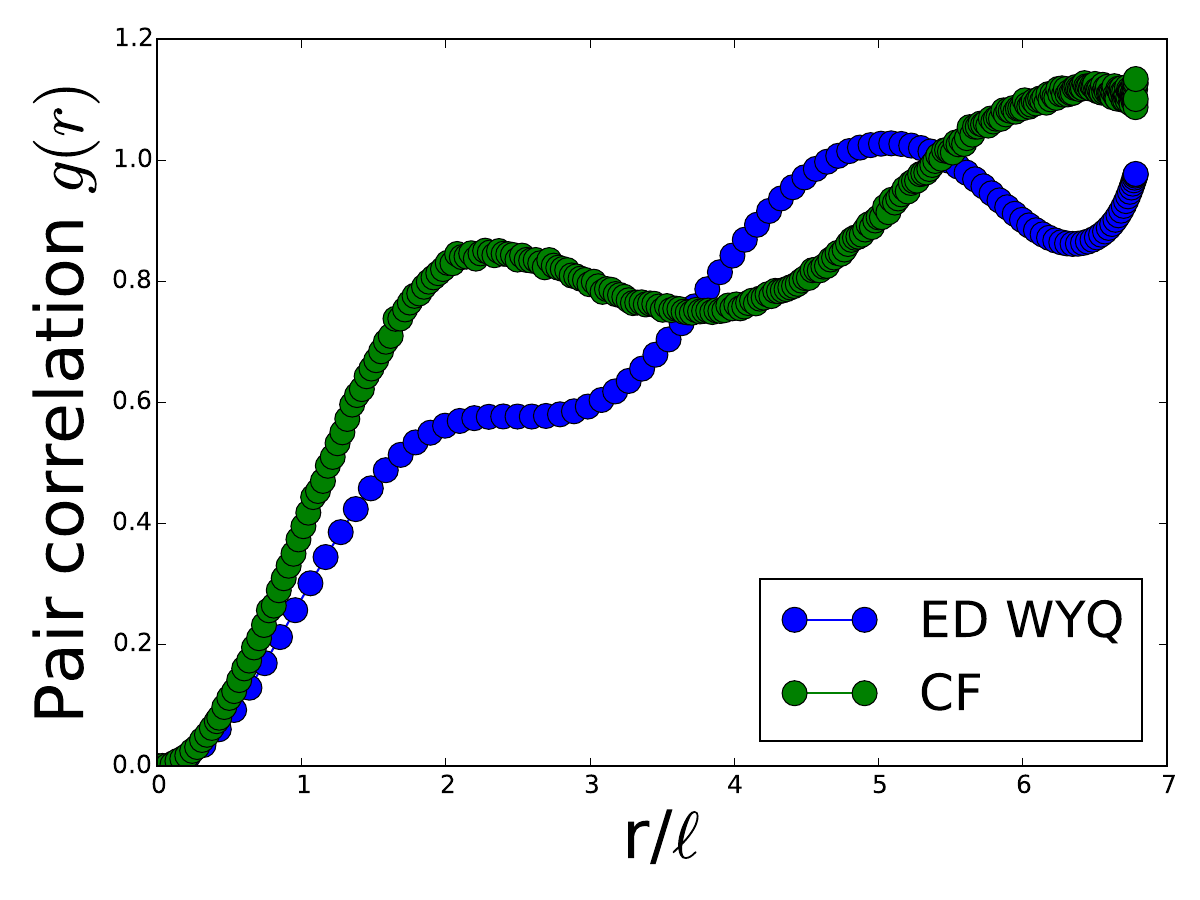}
 \caption{The pair correlation function obtained from the WYQ state by exact 
diagonalization of the hollow-core model with $N_e=10$ fermions is displayed in blue. A composite 
fermion trial wavefunction with the correct shift and filling factor gives a 
very different type of correlations~: this is the green curve. Both calculations are performed 
on the sphere geometry.}
 \label{pair}
\end{figure}

We have also tried to construct an explicit wavefunction with the WYQ shift.
To do so, one has to 
remove Jastrow-type factors out of the Laughlin wavefunction changing the shift 
but without changing the total filling factor. A way to do this can be found in 
the 
CF construction of wavefunctions. In this theory a composite fermion is a bound 
state of an electron and two quantized vortices. The vortex attachment reduces 
the flux felt by the electron and we have $N^*_\phi=N_\phi-2(N_e-1)$
on the sphere. The particles now occupy effective 
Landau levels called $\Lambda$LLs and not simply the LLL because they feel this
reduced effective magnetic flux. Filling an integer number $p$ of these 
$\Lambda$LLs
leads to the FQHE states at electron filling factor $\nu=p/(2p+1)$.
Implicit in this reasoning is the minimization of some kind of mean-field energy
given by the effective cyclotron energy governing the spacing between the 
$\Lambda$LLs.
If we relax this mean-field type of reasoning and just 
consider the algebraic machinery alone it is possible to fill only an excited 
$\Lambda$ level with CFs and to leave empty the lower-lying $\Lambda$LLs.
Certainly
this is not energetically favorable when using the Coulomb interaction.
However it is not immediately clear what happens with the hollow-core 
$\mathcal{H}_3$
interaction.
This procedure of filling only one higher-lying $\Lambda$LL indeed changes the 
shift but not the filling factor. If we fill only
the second $\Lambda$LL one has a state with shift 5 and filling only the third 
$\Lambda$LL gives the WYQ shift of 7. Such states are by construction orbital 
singlets. 
So we consider a trial wavefunction~:
\be
\Psi_t=\mathcal{P}_{\mathrm{LLL}}\Phi_2 J^2,
\label{TCF}
\ee
where the Jastrow correlation factor $J$ is the Vandermonde determinant on the 
sphere~:
\be
J=\prod_{i<j}(u_i v_i - u_j v_i).
\ee
In this equation $\Phi_2$ is a Slater determinant for the $n=2$ $\Lambda$LL
only and $\mathcal{P}_{\mathrm{LLL}}$ is the projection operator onto the LLL.
To perform this projection in an efficient way we have used the technique 
introduced by Jain and Kamilla~\cite{Jain97,Jain97b}.
By construction this state is an orbital singlet with the WYQ relation between 
flux and number of particles.
We have computed the pair correlation function of this state~:
\begin{equation}
g(\vec{r})=
\frac{1}{\rho N_e}\langle
\sum_{i\neq j}\delta^{(2)}(\vec{r}-\vec{r}_i+\vec{r}_j)
\rangle
\label{paircorr-def}
\end{equation}
where $\rho$ is the density.  
It may be evaluated by Monte-Carlo sampling. The result is given by the green curve 
in Fig.~(\ref{pair}). The same pair correlation function for the WYQ 
state obtained 
from direct exact diagonalization is given by the blue curve in the same figure.
While they both have a complex structure they are very different. So we 
conclude that it is 
unlikely that the CF wavefunction Eq.~(\ref{TCF}) can be used to describe the WYQ state
at $\nu=1/3$. Of course this may not exhaust all possibilities of the CF construction.


\subsection{Torus study}
\label{ssec:torus_study}

The geometry used in ED calculations introduces a bias on the states that can be studied.
Notably states with broken space symmetries are frustrated on the sphere and are 
revealed more
clearly on the torus. This is known to be the case for the stripe states that 
appear for half-filling
in the N=2 Landau level and also for the bubble phase for quarter-filling of N=2 
also~\cite{Haldane2000,Rezayi99,Yang2001}.
They are identified by a set of quasi-degenerate ground states that form a one-dimensional
lattice in momentum space for stripe phases or a 2D lattice for bubble phases. 

\begin{figure}[t]
\centering
 \includegraphics[width=0.5\columnwidth]{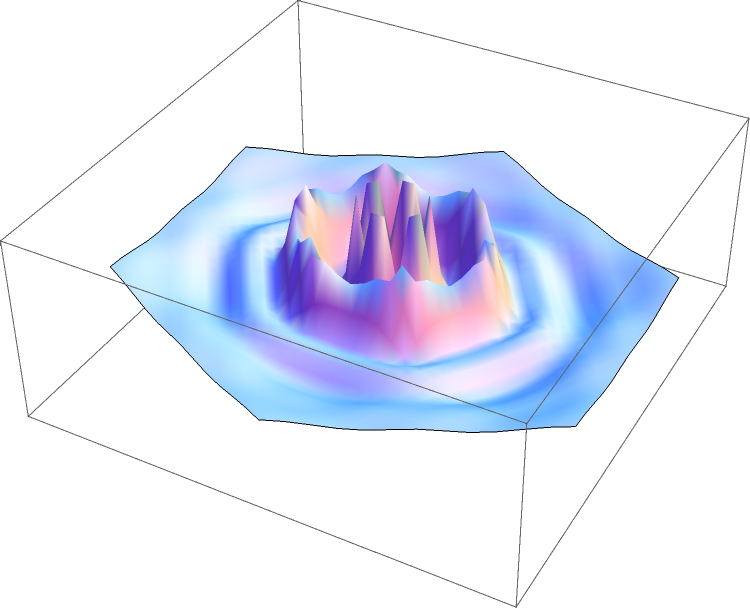}
 \caption{The projected structure factor $S_0\left[\mathbf{q}\right]$ drawn
above the basal plane $(q_x,q_y)$ for $N_e=12$ electrons. The unit cell is 
hexagonal and the ground state is the WYQ state at filling factor 1/3.
The correlations have a double ring structure with a modulation of 
sixfold symmetry due to the choice of the unit cell.}
 \label{hexa}
\end{figure}

We have performed exact diagonalizations on the torus geometry using the 
algebra of magnetic translations which allows us to factor out the overall 
translation invariance. Eigenstates can be classified by two conserved quantum 
numbers $s,t=0,\dots, N_0$ where $N_0$ is the greatest common divisor (GCD) of $N_e$ and $N_\phi$. They 
correspond to the two-dimensional momenta~\cite{Haldane85,HaldaneBook}.
We have performed ED studies on the torus up to $N_e=12$ electrons for the 
$\mathcal{H}_3$ model. In the case of a rectangular unit cell by varying the 
aspect ratio $a_0$ it is possible to favor states breaking translation 
invariance.  For $0.3\leq a_0\leq 1$ there is no evidence for quasi-degenerate 
states. The ground state remains at $\mathbf{K}=0$ and as in the sphere case
there is no well-defined collective excitation mode before reaching a 
higher-lying continuum of excited states.
In addition to spectral
signature an important diagnostic quantity  is 
the 
LLL-projected static structure factor $S_0(\mathbf{q})$ which can be defined 
through the guiding 
center coordinates~$\mathbf{R}_i$~:
\be
S_0\left[\mathbf{q}\right]=
\frac{1}{N_e}\sum_{i\neq j}
\langle \exp{i\mathbf{q}(\mathbf{R}_i-\mathbf{R}_j ) } \rangle .
\ee
When evaluated for the stripe or bubble phases it
has well-defined peaks in reciprocal space
corresponding to the ordering wavevectors.
The sensitivity to changes in the shape of the unit is also an indication that
the state is compressible. This is what we observe in the case of the WYQ state.
The projected structure factor computed in the highly symmetric hexagonal cell
is given in Fig.~(\ref{hexa}). It has a prominent two-ring structure and these rings
are not circular, they are sensitive to the boundary conditions and modulated with the
symmetry of the unit cell
This is 
very different from the Laughlin state which has strongly damped oscillations 
beyond a single central ring surrounding
the correlation hole and this ring is insensitive to the shape of the unit cell.

\begin{figure}[t]
\centering
 \includegraphics[width=0.5\columnwidth]{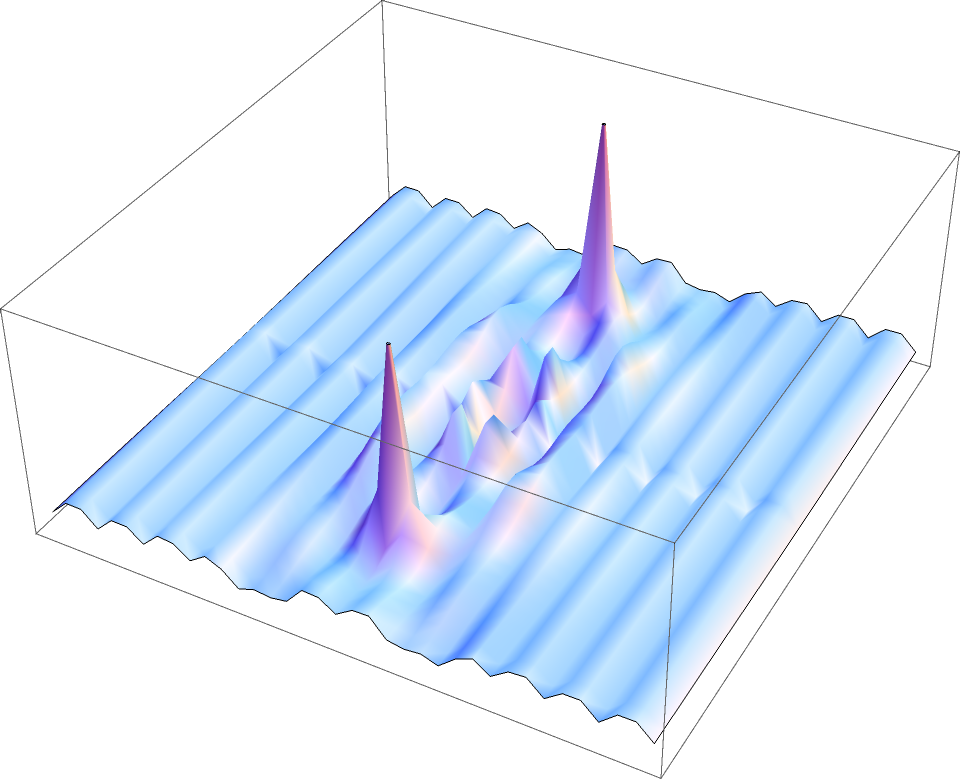}
 \caption{The projected structure factor $S_0\left[\mathbf{q}\right]$ in momentum space in a rectangular unit cell with 
aspect ratio $a_0=0.4$ computed
for $N_e=12$ electrons. There are two sharp peaks suggestive of one-dimensional ordering
as confirmed by the real-space calculation of the pair correlation function.}
 \label{rectangle}
\end{figure}

\begin{figure}
\centering
 \includegraphics[width=0.3\columnwidth]{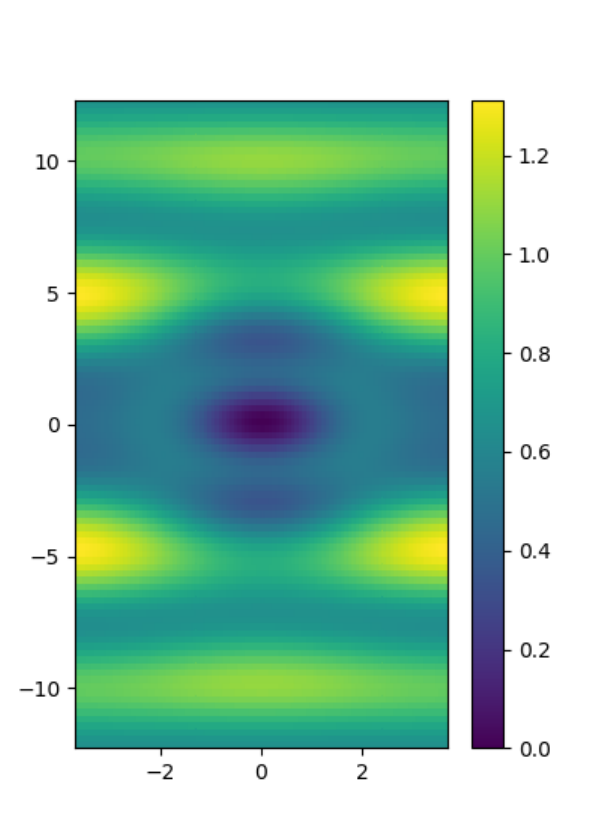}
 \caption{The pair correlation in real space in a rectangular unit cell with 
aspect ratio $a_0=0.3$ computed
for $N_e=10$ electrons. The reference electron is at the center of the rectangle
where it is surrounded by a correlation hole.
There are four stripes that are elongated following the smallest dimension of the unit 
cell.
These stripes are separated by a length $\approx 5\ell$.}
 \label{pairrectangle}
\end{figure}

If we distort the cell to a rectangle with aspect 
ratio 0.4 we find that there are now two well-defined peaks hinting at some
form of one-dimensional ordering~: see Fig.~(\ref{rectangle}). They persist
for $0.3\lesssim a_0\lesssim0.5$. To confirm the spatial pattern we have evaluated the pair
correlation function Eq.~(\ref{paircorr-def}) on the torus geometry. We find that there is
a clear pattern that appears.  A sample calculation is shown in Fig.~(\ref{pairrectangle})

These findings are consistent with a compressible stripe state as the ground 
state of the WYQ model for filling $\nu=1/3$. This identification would be 
complete with the observation of an associated manifold of quasi-degenerate
states. Stripe states~\cite{Koulakov96,Fogler96,Fogler97,Moessner96} have been 
proposed as solutions of the Hartree-Fock approximation for half-filled Landau 
levels with Coulomb interactions with Landau level index at least 2. The 
characteristic wavevector of the stripe then decreases with the LL index.
Since we do not observe the expected manifold of quasidegenerate states associated
with spontaneous breakdown of translation symmetry, it may be that the nature of
the ground state is more subtle.
Since the number of electrons we consider is quite limited
 we turn to the DMRG method for the study of larger systems. 
 This study shows that for large enough systems the ground state is indeed
 a 2-electron bubble phase.


\section{DMRG study in the cylinder geometry}
\label{dmrgres}
\subsection{Hilbert space}

The Hilbert space is the Fock space associated with the lowest Landau orbitals on the cylinder of perimeter $L$, which
are (one-body) wave functions given by~:

\begin{equation}
\phi_n(x,y)=\left(\frac{1}{L \ell \sqrt{\pi}}\right)^{\frac{1}{2}} \exp\left(-\frac{1}{2\ell^2}(x-x_n)^2\right)   \exp\left(i k_n y\right),\quad
 k_n=\frac{2\pi n}{L},\quad x_n/\ell= - \frac{2\pi n}{L/\ell},
\end{equation}
 where the integer $n$ (positive or negative) labels at the same time the angular momentum $k_n$ in the $y$ direction,
and the coordinate $x_n$ of the center of mass of the orbital.

In the following we consider finite cylinders obtained by considering a finite number $N_{\rm orb}=N_{\phi}+1$ of orbitals.
The model does not therefore have a sharp boundary in real space.
Since the spacing in the $x$ direction between two consecutive orbitals is $\delta x=2\pi \ell^2/L\to 2\pi/L$, the length of the cylinder
is of order $L_y\simeq N_{\rm orb} \delta x = 2\pi N_{\rm orb} \ell^2/L \to 2\pi N_{\rm orb} /L$.

For an odd number $N_{\rm orb}$ of orbitals we restrict the index $n$ to be in
the range
$-(N_{\rm orb}-1)/2 \leq n \leq (N_{\rm orb}-1)/2$, and for even $N_{\rm orb}$ we take 
$-N_{\rm orb}/2+1 \leq n \leq N_{\rm orb}/2$. In the following we will denote by $\mathcal{I}$
this set of integer.
The second-quantized form of a generic two-body Hamiltonian reads
\begin{equation}
 \mathcal{H}= \frac{1}{2}\sum_{n_1,n_2,n_3,n_4} \mathcal{A}_{n_1,n_2,n_3,n_4} c_{n_1}^\dag c_{n_2}^\dag c_{n_3} c_{n_4}
\end{equation}
where the matrix elements are related to the real space potential $V(r)$ through
\begin{equation}
 \mathcal{A}_{n_1,n_2,n_3,n_4}= \int d{\bf r}_1 d{\bf r}_2 \phi_{n_1}( {\bf r}_1)^*\phi_{n_2}( {\bf r}_2)^* 
  V({\bf r}_1-{\bf r}_2) \phi_{n_3}( {\bf r}_2)\phi_{n_4}( {\bf r}_1). 
  \end{equation}
We translate the $V_1$-$V_3$ Hamiltonian in the cylinder geometry~:
\begin{equation}
 \mathcal{H}=\mathcal{H}^{(1)} + \mathcal{H}^{(3)} \label{eq:H}.
\end{equation}
with~:
\begin{equation}
 \mathcal{H}^{(1)}= V_1 \frac{(2\pi)^{5/2}}{L^3}\sum_{b,c,d}
(d^2-c^2)\lambda^{c^2+d^2}c_{b+c}^\dagger c_{b+d}^\dagger c_{b+c+d}c_{b},
\label{eq:HV1}
\end{equation}

\begin{equation}
\mathcal{H}^{(3)} = V_3 \frac{(2\pi)^{5/2}}{L^3}\sum_{b,c,d} 
 \lambda^{c^2+d^2} [
\frac{3}{2}(d^2-c^2)+\left(\frac{2\pi}{L}\right)^2 (c^4-d^4) 
  + \frac{1}{6}\left(\frac{2\pi}{L}\right)^4 (d^6-c^6)
 +\frac{1}{2}\left(\frac{2\pi}{L}\right)^4 (c^4 d^2 - c^2 d^4) ] 
 c_{b+c}^\dagger c_{b+d}^\dagger c_{b+c+d} c_{b} ,
\label{eq:HV3}
\end{equation}
where we have defined
\begin{equation}
\lambda = \exp(-2\pi^2\ell^2/L^2) \label{eq:lambda}
\end{equation}
and $b,c,d$ are integers.
Only the terms where $b$, $b+c$, $b+d$ and $b+c+d$ are in $\mathcal{I}$ are kept.
This formulation of the FQHE has already been studied by exact diagonalization~\cite{Soule1,Soule2,Soule3}. The thermodynamic limit is reached for large number of particles but also by sending $L$ to infinity at the same time. At fixed number of particles the $L\rightarrow 0$ limit, called the thin torus limit as well the ``hoop''
limit $L\rightarrow\infty$ are pathological.

\subsection{DMRG results}

\subsubsection{Orbital densities}
\begin{figure}
\includegraphics[width=0.7\linewidth]{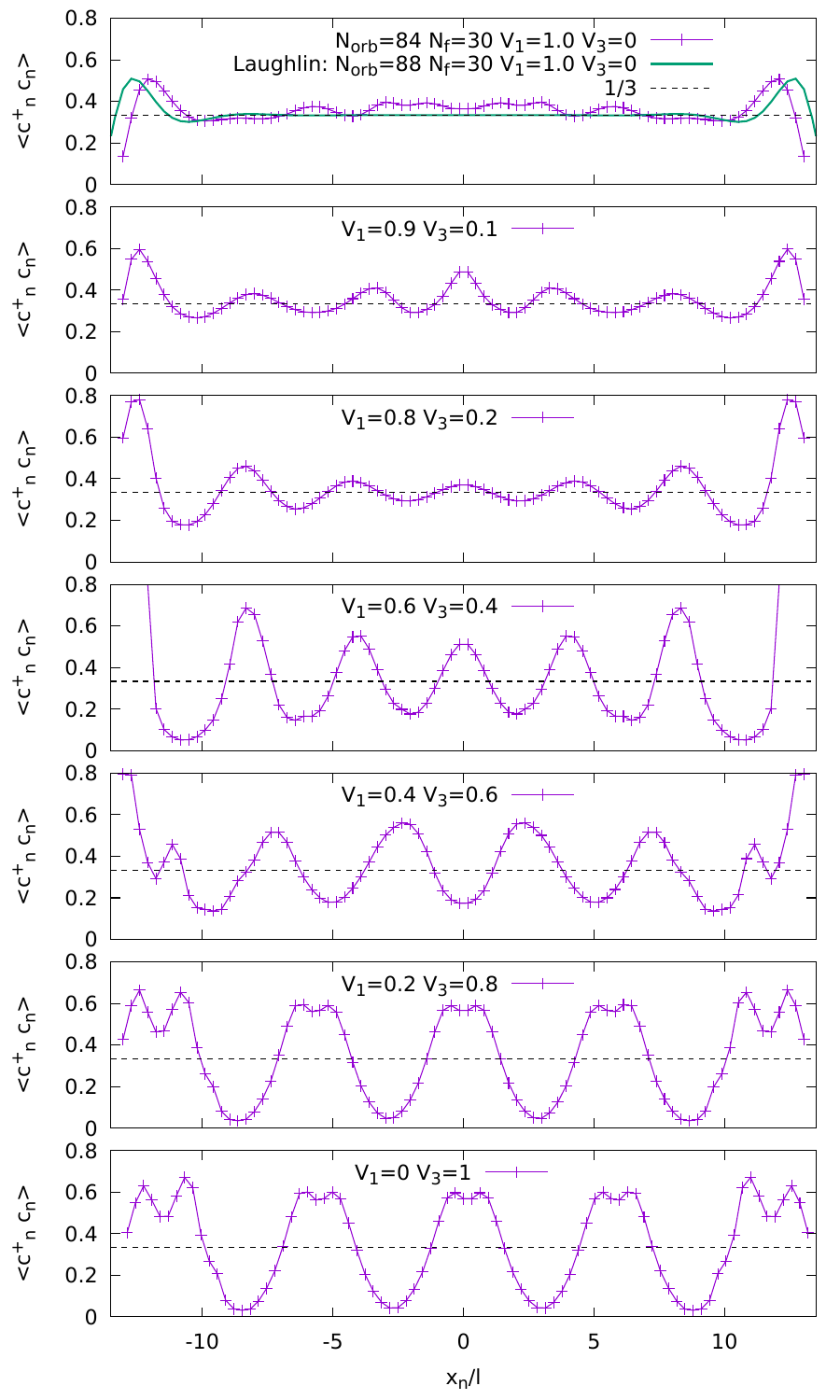}
\caption{Orbital occupancies $\langle c^\dag_n c_n \rangle$ as a function of the (center of mass) 
coordinate $x_n$ of the $n^{\rm th}$ orbital (in units of the magnetic length $\ell$).
$V_1$ varies from 1 to 0 from top to bottom, and $V_3=1-V_1$.
Except for the green curve in the top panel (Laughlin state), all data are obtained for $N_e=30$ 
fermions in $N_{\rm orb}=84$ orbitals ($N_{\rm orb}=3 N_e -6=N_{\phi}+1$).
All calculations performed with $L=20$ and $\chi=8000$.
In the top panel we also plotted the orbital densities for the $\nu=1/3$ zero-energy Laughlin 
state ($N_{\rm orb}=88$). The latter state shows a constant density $n_i\simeq 1/3$ in the bulk 
of the cylinder. 
The pair correlations associated with this series of states are displayed in Fig.~\ref{Fig:pair_correl_scan}. 
} 
\label{fig:scan}
\end{figure}

\begin{figure*}
\includegraphics[width=0.7\linewidth]{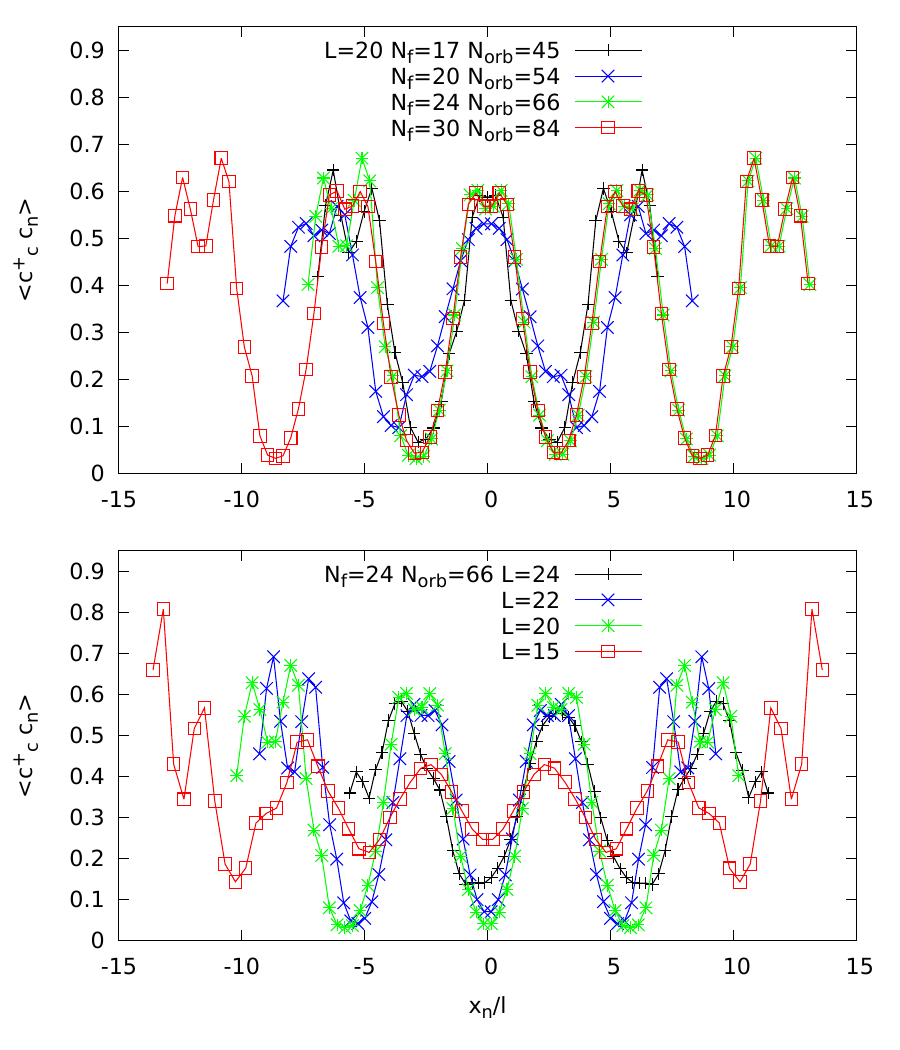}
\caption{Orbital occupancies as a function of the position $x_n$ for $V_3=1$ $V_1=0$ and $N_{\rm orb}=3 N_e -6$, showing some robust large-amplitude modulations with a period $\simeq 5.8 \ell$.
Top: Evolution of the density profile with the system size, varying $N_{\rm orb}$ and the number of fermions $N_e$, keeping
$N_{\rm orb}=3N_e-6=N_{\phi}+1$ fixed and perimeter $L=20$.
The data for $N_e=24$ (with a density {\em minimum} in the center of the cylinder) have been shifted by a half period ($x\to x+2.9\ell$), to allow for a comparison with the other system sizes (which have density {\em maxima} in the center of the system).
Bottom: Density profiles for different values of $L$, with $N_e=24$ and $N_{\rm orb}=66$. 
The data for $L=24$ has been shifted by a half period ($x\to x+2.9\ell$), to allow for a comparison with the other system sizes.
Note that the red curve in the upper panel ($N_e=30$ and $N_{\rm orb}=84$) is the same as the curve in the bottom panel of Fig.~\ref{fig:scan}.} 
\label{fig:orb_dens}
\end{figure*}

The DMRG results are summarized in Fig.~\ref{fig:scan} and \ref{fig:orb_dens}, where the mean orbital 
occupancies $\langle c^\dag_n c_n \rangle$ are displayed.

Fig.~\ref{fig:scan} shows the evolution of the orbital density profile when the Hamiltonian changes 
from the pure $V_1$ model to
the pure $V_3$ model, for $N_e=30$ fermions. At $V_1=1$ and $V_3=0$ we expect the ground-state to be 
an incompressible Laughlin state populated/excited with 4 (fractional) quasi-electrons. This is in 
rough agreement with the density profile displayed in the top panel of
Fig.~\ref{fig:scan} (the homogeneous density of the zero-energy Laughlin state is also displayed for comparison). 

When increasing $V_3$ (and decreasing $V_1=1-V_3$ accordingly) some strong density modulations appear.
The number of density maxima (or bumps) evolves as a function of $V_3$. This is presumably due to the 
fact that the optimal separation between the maxima, which would minimize the energy in an infinite 
system, is evolving continuously with  $V_3$. But the finite size system has to accommodate a finite number of maxima.
For $V_1=0.6$ we can identify 7 maxima (including the two on the edges), while this number is only 5 for the hollow-core model ($V_1=0$). 
This shows that inter-maxima distances grow with $V_3$. From the present data we can estimate this 
distance to be around 5.8 magnetic lengths for the hollow-core Hamiltonian $\mathcal{H}^{(3)}$.

It is important to  check the robustness of these density modulations with respect to the system size. 
The upper panel of Fig.~\ref{fig:orb_dens} displays the orbital occupancies in the pure hollow-core model, 
and varying number of fermions from 17 to 30 (keeping $N_{\rm orb}=3N_e-6=N_{\phi}+1$). The density modulations
appear to be quite robust, showing essentially the same amplitude for 17 and 30 fermions (but a 
slightly reduced amplitude for $N_e=20$). We also have a striking similarity between the data 
for $N_e=30$ and $N_e=24$.
Concerning the dependence on the perimeter $L$ (at fixed $N_e$ and fixed $N_{\rm orb}$), the 
lower panel of Fig.~\ref{fig:orb_dens} shows that the amplitude of the modulation is almost 
the same for $L=20$, 22 and 24. The most elongated cylinder, with $L=15$, shows some reduced 
modulations. All these results strongly suggest that the observed modulations are not a spurious finite size effect.

At this stage these modulations
could be interpreted in two ways i) as a stripe state or ii) as a crystal state (or bubble crystal) state. In the first case
the system would be translation invariant in the $y$ direction, and the stripes would be here rolled 
up in the periodic direction ($y$) of the cylinder. In the second scenario, the state would spontaneously break
the translation symmetry in both directions in the thermodynamic limit. On a finite system this symmetry
necessarily remains unbroken, and the observed stripe-like modulations would come from the fact that the 
finite-size ground-state is the projection of a broken symmetry state in the $\hat J=0$ sector, hence 
translation invariant in the $y$ direction. As we will see below, the analysis of the pair correlations 
allows us to distinguish the two situations and, in the present case, clearly points toward the scenario ii) 
for the $V_3=1$ model.

\subsubsection{Correlations and signatures of a bubble phase}

As discussed in the previous subsection, the orbital occupancies show strong indications of a broken translation symmetry in the $x$ direction.
To detect a possible translation symmetry breaking occurring in the $y$ direction too, we now analyze the
pair correlations $G({\bf r_1},{\bf r}_2)$, as defined in Eq.~(\ref{eq:G_def}).

\begin{figure*}
\includegraphics[width=0.8\linewidth]{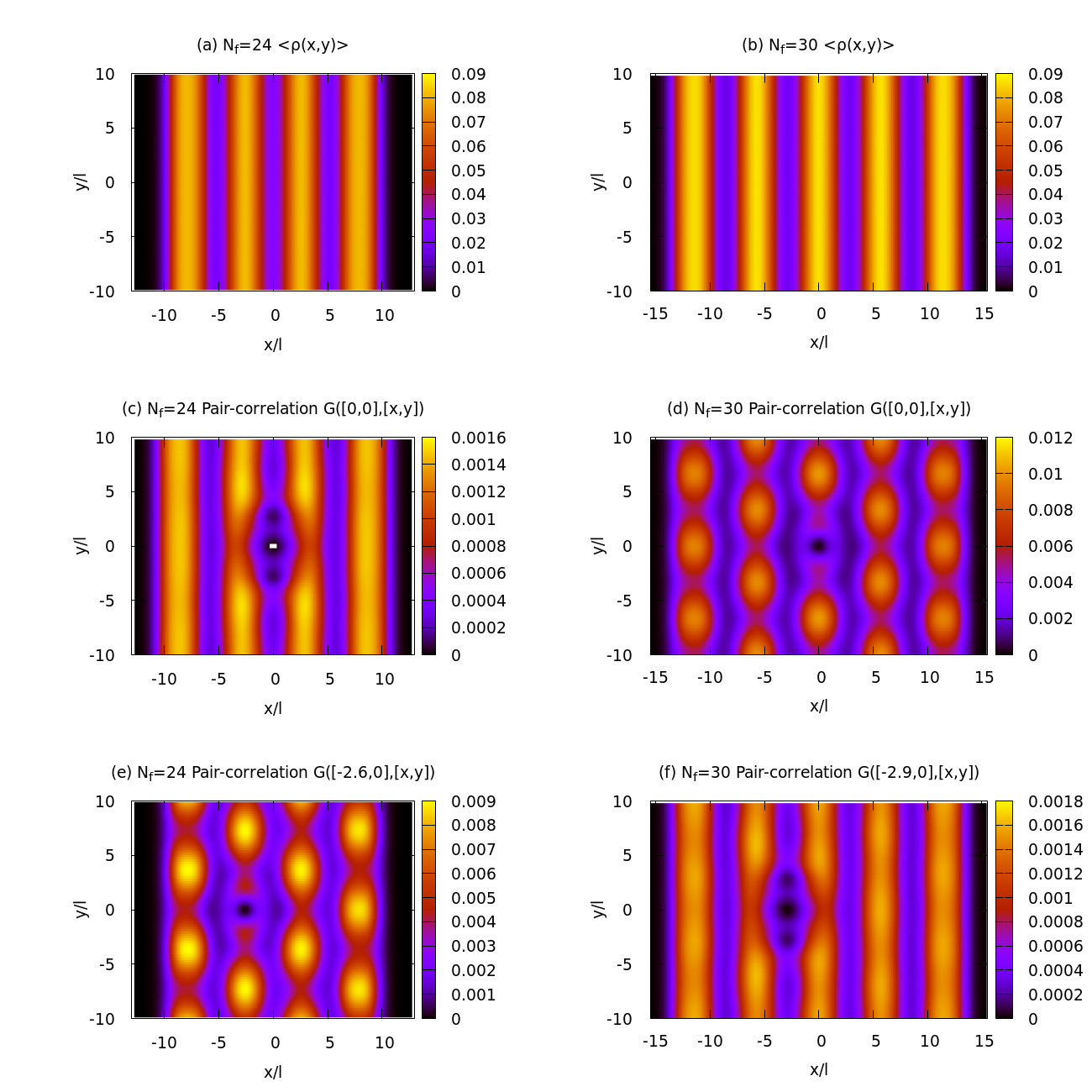}
\caption{Density and pair correlations for the hollow-core $\mathcal{H}^{(3)}$ model.
(a)-(b) Fermion density $\rho({\bf r})$ (see Eq.~\ref{eq:rho_def})
(c)-(d) Pair correlation $G({\bf O},{\bf r})$ (see Eq.~\ref{eq:G_def})
(e)-(e) Pair correlation $G({\bf r}_1,{\bf r})$ with a reference point shifted
to the left.
(a), (c) and (d): $N_e=24$ $N_{\rm orb}=66$ $L=20$
(b), (d) and (f): $N_e=30$ $N_{\rm orb}=84$ $L=20$.}
\label{Fig:pair_correl}
\end{figure*}

The results are displayed in Fig.~(\ref{Fig:pair_correl}).
First, panels (a) and (b) show the fermion density $\rho({\bf r})$ in real space for two different system sizes 
in the pure $V_3$ model and $N_{\rm orb}=3N_e-6$.  These densities are translation invariant in the $y$ direction, 
as they should be in any eigenstate of $\hat J$. The strong modulations in the $x$ directions are simply 
the real-space counterpart of the orbital occupancy modulations described in the previous paragraph.

As for the pair-correlation data, they show some clear triangular pattern of maxima {\em if the 
reference point ${\bf r}_1$ lies on a maximum of the density}. This is the case 
in panels (d) and (e). The number of unit cells of the triangular lattice is 15 in panel (d), with $N_e=30$ fermions. 
It is equal to 12 in panel (e), corresponding to a system with 24 fermions.
These pair-correlation data should be interpreted as signatures of a bubble phase with two electrons per unit cell.
This pattern turns out to be quite robust with the number of fermions and with $L$, and is already visible
on small systems with $\sim 10$ fermions.

\begin{figure*}
\includegraphics[width=0.9\linewidth]{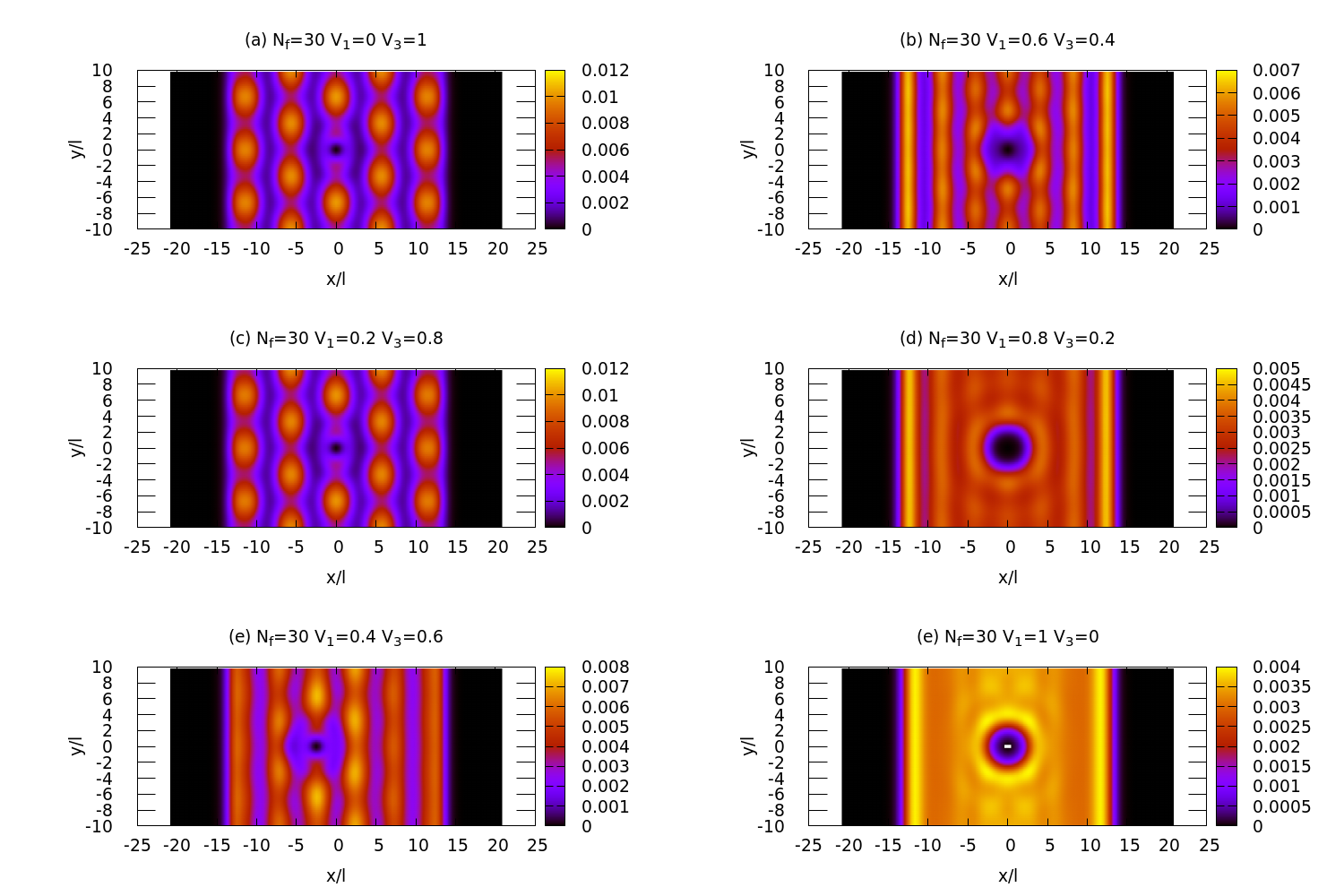}
\caption{Pair correlation for the $\mathcal{H}^{(1)}$--$\mathcal{H}^{(3)}$ model
and different values of $V_1$ and $V_3$ (see the legends above each panel).
It illustrates the evolution from a bubble state in the hollow-core model [panel(a)]
 to the Laughlin liquid [in panel (f)] with short-ranged correlations only and an isotropic correlation hole.
Note that in the panel (e) the reference point of the pair correlation has been chosen slightly 
away from the center of the cylinder  so that it (approximately) matches a maximum of the density.
In all cases $N_e=30$ $N_{\rm orb}=84=N_\phi+1$ and $L=20$, as in Fig.~\ref{fig:scan}.
\label{Fig:pair_correl_scan}
}

\end{figure*}

It is also interesting to look at the evolution of the pair correlations as a function of the parameters $V_1$ and $V_3$
when the Hamiltonian goes from the pure hollow-core model ($V_1=0$ and $V_3=1$) to the hard-core one 
($V_1=1$ and $V_3=0$). This is illustrated in Fig.~\ref{Fig:pair_correl_scan}. Locating precisely the phase 
transition(s) along this path goes beyond the scope of this study, but the present data already indicate that the 
bubble phase should be present at least for $0\leq V_1 /V_3 \leq 0.25$. The nature of the intermediate region 
where $V_3 \simeq V_1$ would deserve some further study, but the
pair correlations displayed in panels (b) and (e) suggest the possibility of a stripe phase which, 
contrary to the bubble phase, would be translation invariant in the $y$ direction.

Bubble states have already been found numerically using DMRG in higher Landau levels \cite{shibata_ground-state_2001,yoshioka_dmrg_2002}.
Our conclusion is related to the results of Ref.~\onlinecite{lee_structures_2002}, which concluded (based on 
the Hartree-Fock calculations for composite fermions) that bubble states with 2 fermion per cell are energetically 
favored compared to stripes for the Coulomb interaction and $\nu \lesssim 0.4$.
In Ref.~\onlinecite{regnault2017}, based on exact diagonalization, it was claimed that
the  $V_1-V_3-V_5$ model at $\nu =\frac{1}{3}$ realizes a stripe or a smectic state for low $V_5$ and large $V_3$.
In view of our results, it seems highly plausible that this phase is in fact nothing but a 2-electron bubble state.

Finally, let us comment on panels (c) and (f) of Fig.(8), where no triangular structure can be seen. 
They correspond to cases where the reference point is at a density minimum (see panels (a) and (b)), 
and where the density is in fact close to zero. Forcing a fermion to be there selects particle 
configurations with a low probability in the many-body wave-function. The fact that the resulting correlation 
pattern is stripe-like
can be interpreted by the fact that picking a low-density point for the reference location does not select 
(does not ``pin'') one broken-symmetry state of the bubble crystal (which is obvious in the limit where the density
at the reference points really vanishes).

It is important to note that the pair correlation function we measure indicates a quantum state
that breaks both rotation and translation symmetry as expected for a bubble phase. A nematic
phase would break only rotation symmetry so its signature would be a correlation hole inside
a uniform fluid state but with an asymmetric pattern around this hole, the preferred direction
being fixed by the boundary conditions. This is not what we observe here. Indeed when we are
closer to the pure $V_1$ model the pair correlations have a perfectly circular correlation hole
and when this phase is destroyed by the effect of $V_3$ pseudopotential there is appearance
of a modulation pattern that breaks the translation symmetry in the whole system, well beyond the 
extent of the hole while a nematic state would have an asymmetric hole surrounded by
a uniform background.

\section{Conclusions}
We have studied the hollow-core $\mathcal{H}_3$ model 
with repulsive interactions only in the RAM $m=3$ channel at filling 
fraction $\nu=1/3$. It has been observed by WYQ that many spectral 
signatures of
an incompressible state are present on the sphere geometry when the shift is 
taken to be 7. However the scaling of gaps calculated on the corresponding both 
to neutral and 
charged excitations points to a compressible state in the thermodynamic limit. 
This is in agreement with recent calculations on the torus 
geometry~\cite{regnault2017}. To clarify the nature of this state we have 
computed
the spatial correlations of the WYQ state on the torus by using the projected 
static 
structure factor. For a hexagonal or square unit cell this quantity has a 
modulated
double-ring  structure unlike that of the Laughlin liquid at the same 
filling factor. When tuning the aspect ratio of a rectangular cell in the range
$0.3\lesssim a_0\lesssim 0.5$ the structure factor develops two very sharp peaks
indicating the presence of a one-dimensional ordering pattern as is the case
of half-filled Landau levels of indices $N\geqslant 2$ for the Coulomb 
interaction. This is confirmed by computation of the pair correlation
in real space in this geometry that shows a stripe modulation with a characteristic
length $\approx 5\ell $.

Since these results are severely size-limited we have used the DMRG algorithm
on the cylinder geometry to study much larger systems of up to 30 fermions.
We have focused again on the pair correlation function to understand the nature of the ground state.
While an incompressible ground state would just exhibit a correlation hole
and no prominent structure beyond that, we find a clear appearance of a two-dimensional
arrangement (for large enough systems beyond those available in the torus or sphere geometry) of modulated density and each overdensity contains exactly two electrons.
This is evidence for the 2-electron bubble phase in line with Hartree-Fock
theory\cite{Koulakov96,Fogler96,Fogler97,Moessner96} for Coulomb interaction
in higher Landau levels.

In the pair correlation measurements we find evidence for breakdown of both translation
and rotation symmetry breaking. This is consistent with a bubble phase but not with a nematic phase
that would appear as a uniform state beyond an asymmetric correlation hole.
It is likely that one would need to add extra pseudopotentials like $V_5$ to realize
a nematic state~\cite{BoYang2020}.

The bubble phase is compressible in agreement with our findings on gap scaling
in the sphere geometry. This coherent picture shows that the hollow-core model
is not a good candidate for describing the effective theory of composite fermions
in the second $\Lambda LL$, a remark that was made in Ref.~\onlinecite{Mukherjee2014}.
If an unconventional state describes the physics of CF in this case then our results
imply that one should go beyond the pure hollow-core model to capture an eventual 
incompressible phase with the WYQ shift.


\begin{acknowledgments}
We acknowledge discussions with O. Golinelli, Ph. Lecheminant and J.-G. Luque. 
Acknowledgments are also due to E. H. Rezayi and S. H. Simon for useful correspondence. 
One of us (TJ) thanks J. K. Jain for an interesting discussion.
T. M. was supported by the research grant of the Senshu Research Abroad Program
(2018).
Part of the numerical 
calculations were performed under the project allocation 100383 from 
GENCI-IDRIS-CNRS.
We also acknowledge CEA-DRF for providing us with CPU time on the supercomputer
COBALT at GENCI-CCRT.
\end{acknowledgments}

\appendix

\section{First and second-quantized forms of the density in real space}

In first-quantized form, the mean density particle density is defined
by
\begin{equation}\label{eq:rho_def}
 \rho({\bf r})=\sum_{i=1}^{N_e}\langle \delta({\bf r}-\hat {\bf r}_i)\rangle,
\end{equation}
where the sum runs over all the particles.
In second-quantized form it becomes
\begin{eqnarray}
\rho({\bf r})&=& \langle \hat\psi^\dag ({\bf r}) \hat\psi ({\bf r}) \rangle \\
  &=&\sum_{n,n'=1}^{N_{\rm orb}} \phi_n^*({\bf r}) \phi_{n'}({\bf r})   \langle c^\dag_n c_{n'}\rangle \\
\end{eqnarray}
where
$\hat\psi^\dag ({\bf r}) = \sum_n \phi^*({\bf r})c^\dag_n$ and the sums over $n$ and $n'$ run over all the orbitals (single-particle states).
In cases where the total angular momentum is conserved $\langle c^\dag_n c_{n'}\rangle$ vanishes unless $n=n'$ and
one gets
\begin{eqnarray}
 \rho({\bf r})&=&\sum_{n}^{N_{\rm orb}} | \phi_n({\bf r})|^2  \langle c^\dag_n c_{n}\rangle \\
   &=&\frac{1}{L\ell\sqrt{\pi}}\sum_{n}^{N_{\rm orb}} \exp\left(\frac{(x-x_n)^2}{\ell^2}\right)  \langle c^\dag_n c_{n}\rangle.
\end{eqnarray}
Writing ${\bf r}=(x,y)$, the above density is independent of $y$ and gives Eq.~\ref{eq:rhox}.
The above normalization insures that $\int d^2{\bf r}  \rho({\bf r}) = \sum_{n}^{N_{\rm orb}} \langle c^\dag_n c_{n}\rangle = N_e$, as it should.

\section{First and second-quantized forms of the two-point correlations}

The density-density correlation function $D$ can be defined as follows~:
\begin{equation}
 D({\bf r}_1,{\bf r}_2)=\sum_{i,j}\langle \delta({\bf r}_1-\hat {\bf r}_i) \delta({\bf r}_2-\hat {\bf r}_j)        \rangle,
\end{equation}
where the sum runs over all the particles.
In second-quantized form it becomes~:
\begin{eqnarray}
 D({\bf r}_1,{\bf r}_2) &=& \langle  \hat\psi^\dag ({\bf r}_1) \hat\psi ({\bf r}_1)  \hat\psi^\dag ({\bf r}_2) \hat\psi ({\bf r}_2)   \rangle \\
  &=&
  \sum_{i,j,k,l=1}^{N_{\rm orb}}
    \phi^*_i ({\bf r}_1) \phi_j ({\bf r}_1)  \phi^*_k ({\bf r}_2) \phi_l ({\bf r}_2) \langle  c^\dag_i c_j  c^\dag_k c_l     \rangle.
\end{eqnarray}

Another quantity of interest is the pair correlation function~:
\begin{equation}\label{eq:G_def}
 G({\bf r}_1,{\bf r}_2)=\sum_{i\ne j}\langle \delta({\bf r}_1-\hat {\bf r}_i) \delta({\bf r}_2-\hat {\bf r}_j)        \rangle,
\end{equation}
which, in second quantization gives~:
\begin{eqnarray}
 G({\bf r}_1,{\bf r}_2) &=& \langle  \hat\psi^\dag ({\bf r}_1) \hat\psi^\dag ({\bf r}_2) \hat\psi ({\bf r}_2)   \hat\psi ({\bf r}_1)   \rangle \\
  &=&
  \sum_{i,j,k,l=1}^{N_{\rm orb}}
    \phi^*_i ({\bf r}_1) \phi^*_j ({\bf r}_2) \phi_k ({\bf r}_2)\phi_l ({\bf r}_1) \langle  c^\dag_i c^\dag_j  c_k c_l     \rangle.
\end{eqnarray}

\section{DMRG implementation}
\label{sec:dmrg_impl}

Several authors have employed DMRG simulations to study fractional quantum Hall effect (FQHE) 
problems~\cite{shibata_ground-state_2001, yoshioka_dmrg_2002, shibata_ground_2003, feiguin_density_2008, kovrizhin_density_2010, zhao_fractional_2011, zaletel_infinite_2015, zhu_fractional_2015, zhu_topological_2015, johri_probing_2016, zhu_fractional_2016,zuo_how_2020}.
In the present work we performed some matrix-product-state (MPS) and matrix-product operator (MPO) based finite-size DMRG.
Our implementation is based on the ITensor~\cite{itensor3} library.

\begin{figure}
\includegraphics[width=0.7\linewidth]{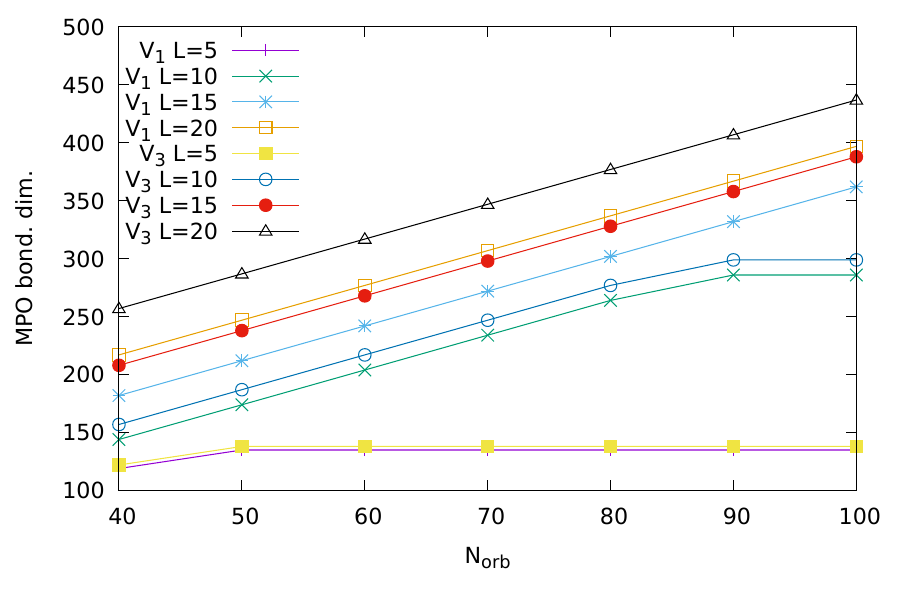}
\caption{MPO bond dimension as a function of the number of orbitals, for the $V_1$ and $V_3$ models 
and different values of the perimeter $L$ of the cylinder.
The MPO truncation parameter was set to $\epsilon=10^{-16}$.} 
\label{fig:MPOdim}
\end{figure}

The Hamiltonian Eqs.~\ref{eq:H} is first converted into a matrix-product 
operator (MPO) \cite{SchollwockAnnPhys11}.
This task is performed using the {\tt AutoMPO} and {\tt toMPO} features of the ITensor library.
Although this conversion to an MPO is not exact, in all our calculations the cut-off parameter was set 
to a very small value, $\epsilon=10^{-16}$, ensuring that this approximation has no significant effect on the results presented here.
Typical values for the bond dimension of the MPO are displayed in Fig.~\ref{fig:MPOdim} for the
$\mathcal{H}^{(1)}$ and $\mathcal{H}^{(3)}$ Hamiltonians.

For cylinders with an elongated aspect ratio (small $L$) the parameter $\lambda$ in Eq.~\ref{eq:lambda} is small 
and the dominant interactions are relatively short-ranged, and the MPO bond dimension saturates with 
the cylinder length (parameterized by $N_{\rm orb}$). This can be observed for $L=5$ in Fig.~\ref{fig:MPOdim}.
On the other hand, for larger $L$, $\lambda$  approaches unity and the interactions decay slowly with 
the orbital separation. In this case the MPO bond dimension increases with $L$ and with $N_{\rm orb}$.
Of course, combining both $V_1$ and $V_3$ interactions increases further the bond dimension. As an 
illustration, for the Hamiltonians used  Fig.~\ref{fig:scan} the bond dimension was found to be between 
372 and 389 (depending on the values of $V_1$ and $V_3$). For consistency the range of integers $c,d$ in the Hamiltonian should be taken of order $L$, meaning an increase of complexity when going to the thermodynamic limit.

The code enforces the conservation of two quantum numbers: i) the total number $N_e$ of fermions,
and ii) the total angular momentum $\hat J$ associated to the periodicity of the cylinder in the $y$ direction.
It reads $\hat J=\frac{2\pi}{L}\sum_{n\in \mathcal{I}} n c^\dag_n c_n$ when $N_{\rm orb}$ is odd.
When $N_{\rm orb}$ is {\em even}, however, it is convenient to redefine 
$\hat J=\frac{2\pi}{L}\sum_{n\in \mathcal{I}} \left(n-\frac{1}{2}\right) c^\dag_n c_n$. 
With this convention the ``central'' momentum sector,  which is the only momentum sector
that is invariant under a reflection with respect to plane $x=0$, is $\hat J=0$ whatever the parity of $N_{\rm orb}$.
All the results discussed here were obtained in this momentum sector.
In a few cases we also looked at other momentum sectors to check that the lowest energy state indeed has $\hat J=0$.

To detect some possible inhomogeneous states, we will be  interested in the real-space fermionic density $\rho(x,y)$. 
An eigenstate of total angular momentum $\hat J$ is  translation invariant in the $y$ direction,
and the above density becomes a function of $x$ only. In such a case 
the density is a convolution of orbital densities $\left\{ \langle c^\dagger_n c_n\rangle\right\}$ with Gaussians:
\begin{equation}\label{eq:rhox}
\rho(x)\sim\sum_n e^{-(x-k_n)^2}\langle c^\dagger_n c_n\rangle.
\end{equation}
For this reason most of the information is contained in the orbital densities. 

\section{DMRG Convergence}
\label{sec:conv}

Unless specified otherwise, all the results presented here were obtained with an MPS bond dimension $\chi=8000$.
The number of sweeps we performed typically goes from  a few tens to one hundred. This number was determined so that 
the total energy variation between two successive sweeps is smaller than $10^{-8}$.

The Laughlin state is the exact zero-energy ground-state of the pure $V_1$ model when $N_{\rm orb}=3N_e-2=N_{\phi}+1$
and this provides a simple way to estimate the precision on the energy.
With 30 fermions (and $N_{\rm orb}=88$) and $L=20$ we find a (variational) energy $E\simeq10^{-8}$
and an MPS truncation error $\mathcal{O}(10^{-10})$ for such a state when $\chi=6000$.

For more complicated states than the Laughlin wave-function, such as the ground of a $V_1$-$V_3$ 
Hamiltonian we
however expect some larger amount of quantum entanglement and a larger truncation error for the 
same MPS dimension. 
For $L=20$ and $\chi$ of the order of a few thousand, the largest MPS truncation error is of the 
order of $10^{-6}$ (see Tab.~\ref{tab:E0}).

\begin{figure}
\includegraphics[width=0.7\linewidth]{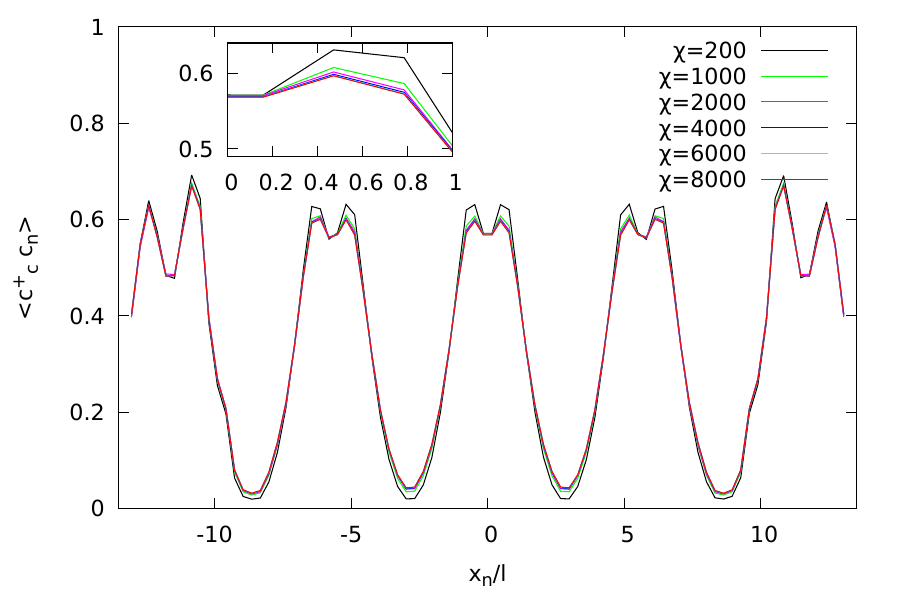}
\caption{Orbital occupancies $\langle c^\dag_n c_n \rangle$ as a function of the 
(center of mass) coordinate $x_n$ of the $n^{\rm th}$ orbital (in units of the magnetic length $l$).
The different curves correspond to different maximal MPS bond dimension $\chi$, from
$\chi=200$ up to $\chi=8000$. The curves for $\chi=2000$, 4000, 6000 and 8000 are almost 
on top of each other at this scale. The inset shows a zoom on a density maximum in the 
center of the system. The associated energies and truncations are given in Tab.~\ref{tab:E0}.
Physical parameters: $N_e=30$, $N_{\rm orb}=84=N_{\phi}+1$ and $L=20$.} 
\label{fig:conv_dens}
\end{figure}

Finally, we also show in Fig.~\ref{fig:conv_dens} the evolution of the orbital densities as a 
function of the MPS bond dimension $\chi$
for a system with 30 fermions, $V_3=1$, $V_1=0$ and $L=20$.
Although the data for $\chi=200$ are clearly not converged enough, the curves for $\chi=1000$, 
2000, 4000, 6000 and 8000 are very close to each other.

\begin{table}
 \begin{tabular}{|c|c|c|}
 \hline
$\chi$	& $E_0$         & $\epsilon$    \\
\hline
200	& 3.7031~	& $7.6.10^{-5}$ \\
1000	& 3.5585~	& $1.7.10^{-5}$ \\
2000	& 3.54003	& $6.5.10^{-6}$ \\
4000	& 3.53068	& $3.5.10^{-6}$ \\
6000	& 3.52796	& $1.4.10^{-6}$ \\
8000	& 3.52705	& $9.5.10^{-7}$ \\
\hline
 \end{tabular}
\caption{Convergence of the energy as a function of the bond dimension $\chi$. The last column,
$\epsilon$, is the largest truncation (discarded weight) during the last sweep.
Parameters of the model: $V_1=0$, $V_3=1$, $N_{\rm orb}=84=N_{\phi}+1$, $N_e=30$ and $L=20$.}
\label{tab:E0}
\end{table}

\section{The $\mathcal{H}^{(3)}$ model series of states at $N_\phi=5N_e-9$}
\label{NewState}

It has been observed in Ref.~(\onlinecite{Wojs2009}) that there is a unique zero-energy ground state for the $V_3$ 
model satisfying $N_\phi=5N_e-9$ starting from $N_e=5$. 
We note that multiplication by a Vandermonde square factor
leads to a state which has exactly the WYQ relation between flux and number of 
particles.
This state is also an orbital singlet as expected for a fluid state without 
ordering.
This feature exists at least till $N_e=11$. Contrary to the case of Laughlin 
wavefunction this peculiar state is not flanked by zero-energy quasiholes
when adding one extra flux quantum. Adding one flux quantum leads to a state
with orbital momentum $L_{\rm orb}=N_e/2$ and a very small but nonzero gap. 
In fact one needs two additional flux quanta
to obtain  new states with zero-energy which are now degenerate.
For an even number of electrons these states are grouped in orbital multiplets
with $L_{\rm orb}=N_e,N_e-2,N_e-4,\dots, 0$ each multiplet appearing exactly once
and there are extra states with $L_{orb}=0$. Apart from the extra singlet states
this is what one expects from two-particle states built with elementary 
quasiholes having $L_{\rm orb}=N_e/2$. When $N_e$ is odd the pattern of states
is identical and the lowest total angular momentum of the set of states is now 
$L_{\rm orb}=1$.

We now focus on the peculiar properties of the unique zero-energy state at 
$N_\phi=5N_e-9$. Notably we find that
the components of the 
ground state eigenvector are all integers after removing the normalization 
factors of the spherical basis and writing the state in terms of the disk 
states Eq.~(\ref{disk}). In fact the property of having integer coefficients is 
true also for small size systems  $N_e=4, N_\phi=9,12$
but the Fock spaces are of very small sizes and it may happen that eigenstates 
are simple. 
On the contrary the states satisfying $N_\phi=5N_e-9$ quickly involve huge 
Fock spaces with growing number of particles and thus the integer decomposition 
is a non-trivial property.
The statement of integer coefficients is quickly limited by the machine 
precision used 
in exact diagonalization. In fact to obtain all integers coefficients one 
has to use quadruple precision already for the state at $N_e=6$ and $N_\phi=21$ 
which
lies in a space of $L_z$ dimension 2137.
In Table (\ref{integers}) we give the first coefficients of the state with 
$N_e=5$. The left column gives the integers while the right column contains
the binary representation of the occupied state in the Slater determinant. 
Since we are dealing with spinless fermions the occupations numbers are only 0 
or 1.

\begin{table}[ht]
\begin{center}
\resizebox{5cm}{!}{
 \begin{tabular}{|r|c|}\hline
    3364    & 11000000100000011 \\
    -13456   & 11000000010000101\\
    22736    & 11000000001001001\\
    25984    & 11000000001000110\\
     -21112  & 11000000000110001\\
    -51968   & 11000000000101010\\
    129920    & 11000000000011100\\
    -13456    & 10100001000000011\\
     35320    & 10100000100000101\\
     -19488   & 10100000010001001\\
     - 22272   & 10100000010000110\\
     -31668    & 10100000001010001\\
     -77952   &  10100000001001010\\ \hline
\end{tabular}
}
\end{center}
\caption{The first 13 coefficients in the expansion of the unique zero-energy 
eigenstate of the $\mathcal{H}^{(3)}$ model for $N_e=5$ and $N_\phi=23$. The left 
column gives the integer while the right column gives the binary representation 
of the occupation numbers of the Slater determinant. There are a total of 252 
integer coefficients in the expansion of the state. The root configuration is 
at the top of the table. These integers are also the coefficients of the 
decomposition of the bosonic state $\mathcal{S}$ onto the Schur basis.}
\label{integers}
\end{table}

Starting from $N_e=6$ particles we find that the polynomial associated with the 
special state has a dominance property i.e. not all possible occupation number 
configuration appear in the expansion. Indeed those with nonzero coefficients 
can be deduced from a root configuration by successive squeezing operations
as happens in many known multivariate special polynomials like the Jack 
polynomials~\cite{BH1,BH2,BH3,BH4,BGS}. The squeezing operation moves a given 
particle from angular momentum $m_1$ to $m_1^\prime$, another particle from
$m_2$ to $m_2^\prime$ with $m_1 < m_1^\prime \leq m_2^\prime < m_2$ and keeping 
the ``center of mass'' intact $m_1+m_2=m_1^\prime+m_2^\prime$. 
The root configuration is $11000000(10000)_k 00000011$ that we note as
$110_6 (10_4)_k 0_6 11$ in a chemistry-like notation. The Laughlin wavefunction
at the same filling factor has also a root configuration but which is 
$(10000)_k 1$. This root is non-trivial only starting from $N_e=6$ because for 
smaller number of particles there are no constraints on the configurations apart
from the $L_z$ angular momentum. We have obtained evidence up to $N_e=11$
a value beyond which the zero coefficients starts to be numerically 
undistinguishable from the nonzero ones.

To discuss the special dominance structure  it is convenient to use also a 
bosonic wavefunction obtained by factoring out a Vandermonde determinant
since the state is antisymmetric~:
\be
P(z_1,\dots,z_N)= \prod_{i<j}(z_i-z_j)\times\mathcal{S}(z_1,\dots,z_N)
\ee
The antisymmetric N-body state is expanded on a Slater determinant basis~:
\be
P(z_1,\dots,z_N)=
\prod_{i<j}(z_i-z_j)\times\mathcal{S}(z_1,\dots,z_N)=
\sum_{\{n_i\}}\mathcal{I}_{\{n_i\}} \,\, \mathrm{det} [\{z_j^{n_i}\}].
\ee
If we divide out both sides by the Vandermonde determinant we see that the 
coefficients $\mathcal{I}_{\{n_i\}} $ determine the expansion of 
the symmetric polynomial $\mathcal{S}$ onto the Schur basis.
 The root partition for the bosonic polynomial $\mathcal{S}$ is given 
by~:
\be
200000100010001\dots 1000002\equiv
20_5(10_3)_k0_52
\ee
This is in fact a partitioning of the total degree of the polynomial.
For $N_e=5$ the root function contains the monomial
$z_1^0 z_2^0 z_3^6 z_4^{12} z_5^{12} $ and the total degree
is $30=12+12+6 $. So we have the following partitions~:
\be
N_e=5:\quad [6, \,\,12,\,\, 12] \equiv 30
\ee
\be
N_e=6:\quad [6,\,\, 10,\,\, 16,\,\, 16]\equiv 48
\ee
\be
N_e=7:\quad [6,\,\, 10,\,\, 14,\,\, 20,\,\, 20]\equiv 70
\ee
\vskip 1cm
The total degree of $\mathcal{S}$ is given by $\frac{1}{2}N_e(4N_e-8)$.
It would be interesting to obtain a closed analytic formula for this special 
state. The numerous known examples~\cite{BGS,BH1,BH2,BH3,BH4}
suggest that this may be possible.
However it is not a \textit{bona fide} Jack polynomial since it is known that 
the rotational invariance of the state constrains both the root partition as 
well as the parameter defining the Jack.

If now we study the $\mathcal{H}^{(3)}$ model on a torus at filling $1/5$ then we 
know 
already that there will be at least one zero-energy state at the center of the 
Brillouin zone $K=0$ which is the non-degenerate Laughlin state. However we 
find more zero-energy states in a complex pattern. For $N_e=5$ we find that 
there is a threefold degenerate state at the center of the Brillouin zone and 
also additional zero-energy states at the zone boundaries~: see Table 
(\ref{tab3})

\begin{table}[h]
\begin{center}
\resizebox{6cm}{!}{
\begin{tabular}{|c|c|c|c |c |}\hline
   $K$   & $(0,0)$ & $(0,N_e/2)$ & $(N_e/2,0)$ & $(N_e/2,N_e/2)$  \\ 
 \hline
  deg.  &  x3    & x1        &  x1       &  x1   \\ \hline
\end{tabular}
}
\end{center}
\caption{The quantum numbers of zero-energy eigenstates for the $\mathcal{H}^{(3)}$ 
model. Here we display the case of $N_e=5$ particles. The two components of the 
wavevector $\mathbf{K}$ are given on the first line in units of $2\pi/L_{x,y}$. 
The calculation has been done in a rectangular unit cell and is insensitive to 
the aspect ratio.}
\label{tab3}
\end{table}

\begin{table}[h]
\begin{center}
\resizebox{6cm}{!}{
\begin{tabular}{|c|c|c|c |c |}\hline
 $K$   & $(0,0)$ & $(0,N_e/2)$ & $(N_e/2,0)$ & $(N_e/2,N_e/2)$  \\ 
 \hline
 deg.  &  x4    & x1        &  x1       &  x1   \\ \hline
\end{tabular}
}
\end{center}
\caption{location of the zero-energy states for $N_e=6$ particles. Same 
definitions as in Table (\ref{tab3}).}
\label{tab4}
\end{table}

\begin{table}[ht]
\begin{center}
\resizebox{12cm}{!}{
\begin{tabular}{|c|c|c|c |c |c|c|c|}\hline
 $K$   & $(0,0)$ & $(0,N_e/2)$ & $(N_e/2,0)$ & $(N_e/2,N_e/2)$ & $(0,N_e/3)$ & $(N_e/3,0)$ &  
$(N_e/3,N_e/3)$\\ \hline
 deg.  &  x7    & x1        &  x1       &  x1            & x1       & x1      
   & x1 \\ \hline
\end{tabular}
}
\end{center}
\caption{location of the zero-energy states for $N=7$ particles. Same 
definitions as in Table (\ref{tab3}). There are now states inside the Brillouin 
zone with zero-energy.}
\label{tab6}
\end{table}

The number of zero-energy states grows with the number of particles in a 
manner reminiscent of the Haffnian state~\cite{Hermanns2011}.
This special wavefunction is related to an irrational conformal field theory
and is presumably gapless~\cite{ReadNonUI,ReadNonUII}. There are several
quantum Hall states including the Haldane-Rezayi spin-singlet 
state~\cite{HR88,Gurarie97,Simon07}, the Gaffnian state~\cite{Simon07}
that share this property. We have been unable to find an explicit analytic
fomula for this special wavefunction. Its very peculiar properties are worth studying.
We have learned that similar findings have been obtained by S. H. Simon in unpublished work.


\end{document}